\documentclass{Interspeech}

\interspeechcameraready

\title{Discl-VC: Disentangled Discrete Tokens and In-Context Learning for Controllable Zero-Shot Voice Conversion \thanks{*Corresponding author}}

\author[affiliation={1}]{Kaidi}{Wang}
\author[affiliation={2}]{Wenhao}{Guan}
\author[affiliation={3}]{Ziyue}{Jiang}
\author[affiliation={1}]{Hukai}{Huang}
\author[affiliation={1}]{Peijie}{Chen}
\author[affiliation={1}]{Weijie}{Wu}
\author[affiliation={1}*]{Qingyang}{Hong}
\author[affiliation={2}*]{Lin}{Li}

\affiliation{School of Informatics}{Xiamen University}{China}
\affiliation{School of Electronic Science and Engineering}{Xiamen University}{China}
\affiliation{}{Zhejiang University}{China}
\email{kaidi@stu.xmu.edu.cn}

\usepackage{comment}
\usepackage{multirow}

\begin{document}

\maketitle

\begin{abstract}
Currently, zero-shot voice conversion systems are capable of synthesizing the voice of unseen speakers. However, most existing approaches struggle to accurately replicate the speaking style of the source speaker or mimic the distinctive speaking style of the target speaker, thereby limiting the controllability of voice conversion. In this work, we propose Discl-VC, a novel voice conversion framework that disentangles content and prosody information from self-supervised speech representations and synthesizes the target speaker's voice through in-context learning with a flow matching transformer. To enable precise control over the prosody of generated speech, we introduce a mask generative transformer that predicts discrete prosody tokens in a non-autoregressive manner based on prompts. Experimental results demonstrate the superior performance of Discl-VC in zero-shot voice conversion and its remarkable accuracy in prosody control for synthesized speech.
\end{abstract}

\section{Introduction}
Voice conversion is the task of transforming the voice of a source speaker into that of a target speaker, while preserving the linguistic content information of the source speaker. 
Zero-shot voice conversion \cite{Li2023SEFVCSE, ma24e_interspeech} generates target speech given the voice of an unseen speaker during training, making the task even more challenging. A widely adopted approach to address this challenge is to decouple linguistic content information and speaker information from the speech. The AutoVC series \cite{pmlr-v97-qian19c,pmlr-v119-qian20a, 9747763}, based on a simple autoencoder architecture, utilizes a carefully designed bottleneck to achieve attribute disentanglement, enabling zero-shot conversions. With the rise of self-supervised models, some researchers have explored disentangling content feature by filtering out timbre information through a bottleneck mechanism within self-supervised representations, such as K-means clustering \cite{Polyak2021SpeechRF}. ContentVec \cite{pmlr-v162-qian22b} directly proposes a new self-supervised representation that effectively filters out speaker information, while minimizing content loss caused by discretization, thereby performing speech conversion using its continuous representations \cite{huang24_interspeech}. Additionally, some studies have introduced generative models such as normalizing flows \cite{casanova2022yourtts} and diffusion models \cite{popov2022diffusionbased,choi2024dddm} into voice conversion, enhancing the quality of the converted speech.

However, in addition to linguistic content information and speaker information, speaking style is also a crucial component in speech \cite{Jiang2023MegaTTSZT,jiang2024megatts, 10889108}. ACE-VC \cite{Hussain2023ACEVCAA} predicts speaking rate and pitch contour through speaker and content embeddings, achieving a controllable and adaptive voice conversion model. Diff-HierVC \cite{choi23d_interspeech} proposes utilizing the diffusion process to generate F0 (fundamental frequency) to achieve more accurate pronunciation and natural intonation in the converted speech. StableVC \cite{Yao2024StableVCSC} introduces a method to explicitly disentangle content, style, and timbre, enabling separate control over the speaking style and timbre of the generated speech. Recently, Vevo \cite{zhang2025vevo} has also introduced a controllable voice conversion model by progressively incorporating style and timbre information using a popular hybrid autoregressive and non-autoregressive architecture. Despite these advancements, there is still room for improvement in the naturalness and prosody similarity of the converted speech in controllable voice conversion.

In this paper, we propose Discl-VC, a controllable zero-shot voice conversion system that explicitly disentangles and separately models the various attributes of speech. Specifically, we decompose speech into three parts: content, prosody, and timbre. We leverage different self-supervised representations and discretization methods to disentangle prosody and content. Given the impressive capabilities of flow matching \cite{lipman2023flow} in the field of audio \cite{NEURIPS2023_2d8911db,Guan2023ReflowTTSAR,Guan2024LAFMAAL,10889258}, we propose a flow matching transformer that leverages in-context learning to achieve fine-grained timbre modeling \cite{10832320, chen-etal-2024-f5tts}. At the same time, we introduce a fully non-autoregressive prosody mask transformer, following the mask-and-predict learning paradigm \cite{chang2022maskgit,ju2024naturalspeech,wang2025maskgct}, which predicts prosody tokens of the generated speech based on a reference speech. Our contributions are summarized as follows:

\begin{figure*}[htbp]
    \includegraphics[width=0.95\linewidth]{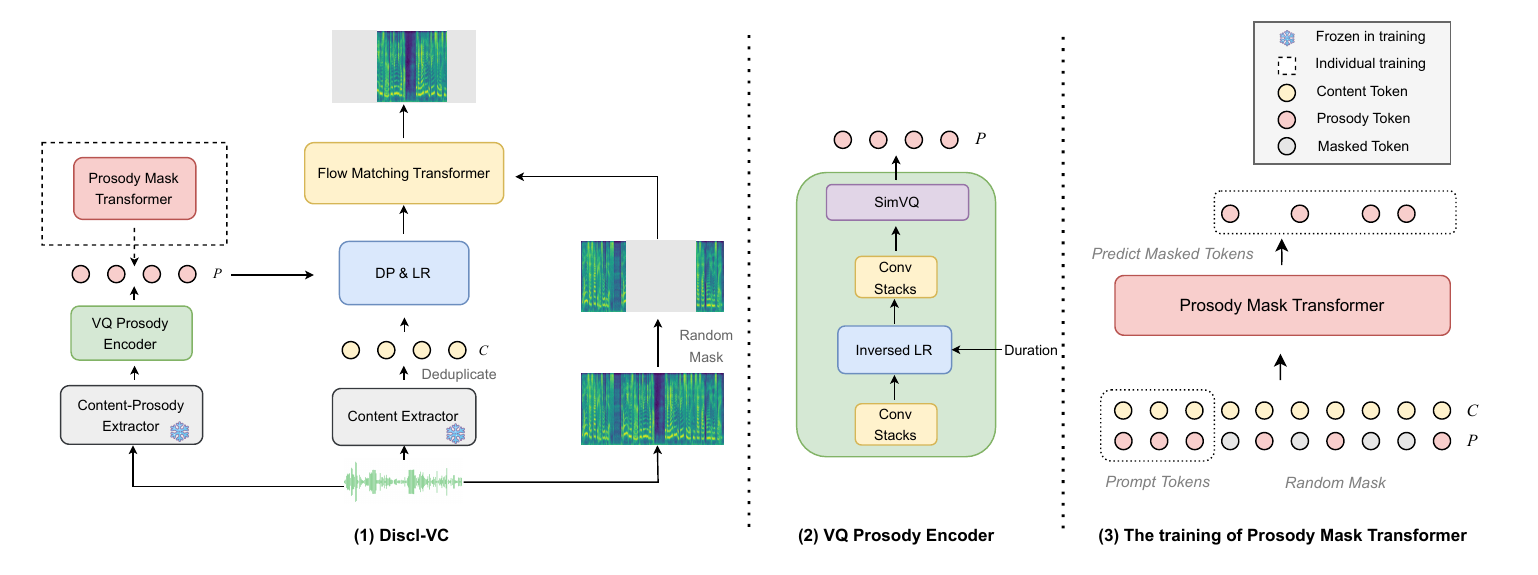}
        \centering
    \caption{The overall architecture of our proposed system.}
    \label{fig:discl-vc}
\end{figure*}

\begin{itemize}
\item We propose a method that disentangles content and prosody information from self-supervised representations, obtaining discrete content tokens and prosody tokens separately to facilitate prosody control.
\item We introduce two non-autoregressive modules: the prosody mask transformer and the flow matching transformer, which leverage in-context learning to perform prosody token prediction and acoustic representation prediction, respectively.
\item Experimental results demonstrate that our model outperforms the baseline in zero-shot voice conversion and prosody related controllable voice conversion tasks, with improved speech quality and more accurate prosody control. Our demos are available at https://wkd12345.github.io/disclvc/.

\end{itemize}

\section{Discl-VC}

\subsection{Speech disentanglement}

The overall architecture of our system is shown in Figure \ref{fig:discl-vc} (1), where the Content Extractor and Content-Prosody Extractor are pre-trained self-supervised models. In this work, we use HuBERT large \cite{9585401} as the Content Extractor and apply K-means clustering on the continuous representations from the 24th layer. The number of clusters is set to 1024. The content tokens obtained from the Content Extractor can be viewed as containing only semantic information while filtering out most timbre and prosody information. We also perform a deduplication process by removing adjacent duplicate tokens, further eliminating duration-related prosody information. This approach allows for the re-prediction of speech token durations, thereby improving the prosody of the generated speech.

For the Content-Prosody Extractor, we directly use ContentVec \cite{pmlr-v162-qian22b}. Since its output continuous representations have already disentangled speaker information and contain almost all of the content and prosody information, this reduces the complexity of the disentangling process. We introduce a Vector Quantization (VQ) Prosody Encoder to extract the prosody information, as shown in Figure \ref{fig:discl-vc} (2), which consists of two convolution stacks, a Inversed length regulator (Inversed LR), and a VQ layer. The Inversed LR adjusts the sequence length based on the duration of the tokens. We propose a VQ layer as a bottleneck to filter content information. We also adopt SimVQ \cite{Zhu2024AddressingRC} to avoid the codebook collapse issue. Specifically, we randomly initialize the codebook vectors and do not update them during training. Instead, we introduce a linear layer \(W\) that applies to the codebook vector \(q\) to generate the quantized result \(z_q\). The vector quantization loss is formulated as follows:
\begin{equation}
\mathcal{L}_{SimVQ} = \lambda \lvert\lvert qW-sg[z]\rvert\rvert^2 + \lvert\lvert z-sg[qW]\rvert\rvert^2
\end{equation}
where \(\lambda\) is the hyper-parameter.

Additionally, since F0 is a crucial component of prosody, we introduce a \(\mathcal{L}_{F0}\) to effectively supervise the model's extraction of prosody information. Specifically, we compute the mean and variance for each speaker and apply Z-score normalization to the ground truth F0 values to obtain speaker-independent prosody. We then use Smooth L1 loss to compute the difference between the predicted and ground truth values \cite{ju2024naturalspeech}. Subsequently, we introduce a duration predictor and a length regulator (DP \& LR) to expand the token lengths based on the predicted durations. The structure of these components follows that of \cite{ren2021fastspeech}.

\subsection{In-context learning modeling}
\subsubsection{Flow matching transformer}

Flow Matching \cite{lipman2023flow} is a generative approach that learns to map a simple distribution to a target data distribution  by learning an ordinary differential equation (ODE). 
The core idea is to define a time-dependent vector field 
\(v_t(x;\theta)\) that gradually transforms the initial distribution into the data distribution, with the objective of minimizing the discrepancy between the true vector field 
\(u_t(x)\) and the learned vector field \(v_t(x;\theta)\).

In this work, we adopt the optimal transport flow-matching objective \cite{lipman2023flow}, where the flow between the initial and target distributions is modeled as a straight line, leading to the OT-CFM loss:

\begin{equation}
\begin{aligned}
\mathcal{L}_{OT-CFM}(\theta) = 
\mathbb{E}_{t, q(x_1),p(x_0)} \Big[\lvert\lvert v_t((1-t)x_0+tx_1)\\- (x_1-x_0)\rvert\rvert^2 \Big]
\end{aligned}
\end{equation}

VoiceBox \cite{NEURIPS2023_2d8911db} was the first to introduce integrating flow matching with transformer for the text-guided speech-infilling task. In this work, we also adopt the mask-and-predict approach, using in-context learning to generate the masked acoustic features. Specifically, during training, we sample a time step \(t\), add a certain level of Gaussian noise to the real mel spectrogram to obtain a noisy version, and then feed this noisy mel spectrogram, along with the input prosody and content features and the masked mel spectrogram, into the flow matching transformer. This module uses the complete prosody and content information, along with the surrounding mel spectrograms, to predict the masked mel spectrogram. During training, we focus on optimizing only the masked portion. For the flow matching transformer, we use DiT blocks with the adaLN-zero structure \cite{chen-etal-2024-f5tts}. We apply classifier-free guidance to generate higher quality speech samples, specifically by setting a 0.2 probability to remove prosody and content tokens, as well as the masked mel spectrogram.

\subsubsection{Prosody mask transformer}

In controllable voice conversion, we need to mimic the speaking style of a reference audio. To achieve this, we introduce a non-autoregressive prosody mask transformer, which predicts the prosody tokens of the source speech based on the reference audio, thereby generating speech with the same semantic content but different speaking styles.

This masked token modeling approach was first introduced in image synthesis \cite{chang2022maskgit} and has since been widely applied to audio fields \cite{Borsos2023SoundStormEP, wang2025maskgct}. Unlike autoregressive models that generate tokens sequentially, this method predicts all tokens in parallel and iteratively refines low-confidence outputs. The training process of the module is shown in Figure \ref{fig:discl-vc} (3), similar to that in \cite{wang2025maskgct}. Specifically, given a sequence of prosody tokens, we introduce a special token and replace selected tokens with this token according to a sine schedule. The module predicts the masked tokens based on the full content token sequence and the unmasked prosody tokens. The module optimizes the masked portions using cross-entropy loss. In the inference phase, the module uses the full reference audio's prompt tokens and the source speech's content tokens, along with completely masked prosody tokens, to iteratively unmask the prosody tokens. At each step, low-confidence tokens are re-masked based on time step until all prosody tokens are generated. The module architecture is consistent with the flow matching transformer. We also apply classifier-free guidance with a 0.2 probability of removing the conditions.

\subsection{Model training}

Our model requires a two-stage training process. First, we jointly train the VQ prosody encoder, DP\&LR, and flow matching transformer. Through this training, the model learns to decouple speech and is capable of performing high-quality zero-shot voice conversion tasks. The loss function for this stage is:
\begin{equation}
\mathcal{L}_{stage1} = 
\mathcal{L}_{Dur} + \mathcal{L}_{SimVQ} + \mathcal{L}_{FMT} + \mathcal{L}_{F0}
\end{equation}
where \(\mathcal{L}_{Dur}\) represents the MSE loss for duration, and \(\mathcal{L}_{FMT}\) represents the flow matching loss.

Afterwards, the trained VQ prosody encoder is used to extract the ground true prosody tokens. These tokens, along with the content tokens, are then used to continue training the prosody mask transformer. The training objective is:
\begin{equation}
\mathcal{L}_{stage2} = \mathcal{L}_{PMT}
\end{equation}
where \(\mathcal{L}_{PMT}\) represents the cross-entropy loss for prosody mask transformer.
After training, the module is capable of predicting prosody tokens based on reference audio, enabling controllable voice conversion.

\section{Experiments}

\subsection{Experiments setup}
\subsubsection{dataset}
We use the Librilight small+medium dataset for training, which contains approximately 6k hours of English speech data. During training, we only use speech with a duration between 4 to 20 seconds. We evaluate the model's performance on voice conversion tasks using the VCTK and ESD datasets. Specifically, we utilize the VCTK dataset to perform zero-shot voice conversion tasks. We randomly select 5 male and 5 female speakers, with 5 fixed audio samples from each speaker, totaling 50 audio samples. This results in 450 test speech pairs. To further evaluate the model's capability in controlling prosody, we additionally randomly select two male and two female speakers from the ESD English dataset, using one audio sample for each emotion, totaling 20 audio samples as prosodic prompts. These are used to conduct voice conversion tasks under both prosody preserved and prosody converted scenarios. For prosody-preserved voice conversion, we use the ESD audio as the source and the VCTK audio as the target, while for the prosody-converted voice conversion, the roles are reversed. Each of these two tasks contains 1,000 test audio samples. It is important to note that all the audio used for testing was unseen during training.

\subsubsection{Implement details}
We use the pre-trained RMVPE\footnote{https://github.com/Dream-High/RMVPE} to extract F0. For the VQ prosody encoder, we set the kernel size of the convolution layers in the conv stacks to 3, the hidden size to 384, the size of the codebook in VQ to 2048, and the codebook vector dimension to 256, in line with \cite{Jiang2023MegaTTSZT}.
 The duration predictor is made up of two convolution layers with a kernel size of 3, a hidden size of 384, and a dropout rate of 0.5. The transformers both consist of 12 layers, each with 12 attention heads. We also apply a rotary position embedding (RoPE) for self-attention. The embedding dimension is 768, and the feed-forward network dimensions are set to 1536 for the flow matching transformer and 3072 for the prosody mask transformer.

We downsample the audio to 16 kHz and convert it into 80-dim Mel spectrograms with a window size of 1280 and a hop size of 320. We train the model using two 4090 GPUs, with a batch size of 12,000 speech frames per GPU. For the first phase of training, the model is trained for 500k steps, while the second phase, the prosody mask transformer, is trained for 300k steps. We use the AdamW optimizer with \(\beta_1\)=0.9, \(\beta_2\)=0.98, and an initial learning rate of 2e-4, which then linearly decays to 5e-5 over 600k steps. 
During inference, the number of inference steps for both transformers is set to 16. The guidance scale for the flow matching transformer is set to 1. For the prosody mask transformer, the guidance scale is set to 2.5. We use top-k sampling with k set to 40, and the sampling temperature anneals from 1.5 to 0. We add Gumbel noise to the token confidences during the remasking process. A pre-trained BigVGAN \cite{lee2023bigvgan} vocoder is used to generate the final waveform.

\begin{table}[htb]
  \caption{Zero-shot voice conversion results.}
  \label{tab:vctk}
  \centering
  \setlength{\tabcolsep}{1.2mm}{
    \scalebox{0.7}{
  \begin{tabular}{lcccccc}
    \toprule
    \textbf{} & \textbf{N-MOS} ($\uparrow$) & \textbf{UTMOS} ($\uparrow$)
    & \textbf{WER} ($\downarrow$) & \textbf{F0\_Corr} ($\uparrow$) & \textbf{SECS} ($\uparrow$) & \textbf{\#Params} \\
    \midrule
    FAcodec & 3.44 ± 0.10 & 3.580 & 1.341 & 0.968 & 0.892 & 138M \\
    Vevo  & 4.26 ± 0.07 & 3.970 & 2.705 & 0.966 & 0.930 & 922M \\
    \midrule
    Discl-VC & 4.31 ± 0.07 & 4.079 & 1.946 & 0.973 & 0.929 & 131M \\
    \bottomrule
  \end{tabular}
  }}
\end{table}

\subsubsection{Baseline and metrics}
We compare our model with FAcodec\footnote{https://huggingface.co/amphion/naturalspeech3\_facodec} and Vevo\footnote{https://huggingface.co/amphion/Vevo}, both of which use the officially released pre-trained models. We conducted both objective and subjective evaluation experiments.
For objective metrics, we used UTMOS\footnote{https://github.com/sarulab-speech/UTMOS22} to assess the speech quality, employed the Whisper-medium.en\footnote{https://github.com/openai/whisper} to evaluate the speech's WER (Word Error Rate), and utilized the pre-trained WavLM\footnote{https://huggingface.co/microsoft/wavlm-base-plus-sv} model to extract speaker embeddings and compute the Speaker Encoder Cosine Similarity (SECS). Additionally, to evaluate the accuracy of prosody control, we introduced pitch-related metrics such as the F0 Pearson correlation (F0\_Corr).
For subjective evaluation, we use N-MOS and P-MOS to evaluate the naturalness and prosody similarity of the generated samples, respectively. We randomly selected 30 samples for each task and invited 10 listeners to rate the audio. 

\subsection{Main results}

For the zero-shot voice conversion task, the evaluation results are shown in Table \ref{tab:vctk}. Our model achieved the best speech quality, prosody similarity, and excellent timbre reproduction with fewer parameters, while the WER performance falls between the two baselines, demonstrating the superiority of our model for zero-shot voice conversion. 

\begin{table}[htb]
  \caption{Results of objective metrics for prosody related zero-shot voice conversion.}
  \label{tab:esd}
  \centering
    \setlength{\tabcolsep}{1.2mm}{
  \scalebox{0.7}{
  \begin{tabular}{llccccc}
    \toprule
    & \textbf{} & \textbf{UTMOS} ($\uparrow$)
    & \textbf{WER} ($\downarrow$) & \textbf{F0\_Corr} ($\uparrow$) & \textbf{SECS} ($\uparrow$) & \textbf{\#Params} \\
    \midrule
    \multirow{3}{4em}{Prosody Preserved}
    & FAcodec & 3.377 & 4.336 & 0.942 & 0.777 & 138M \\
    & Vevo  & 4.057 & 2.461 & 0.928 & 0.838 & 922M \\
    \cmidrule{2-7}
    & Discl-VC & 4.119 & 3.171 & 0.941 & 0.839 & 131M \\
    \midrule
    \multirow{2}{4em}{Prosody Converted}
    & Vevo & 3.783 & 7.720 & 0.799 & 0.892 & 922M \\
    \cmidrule{2-7}
    & Discl-VC & 4.063 & 4.139 & 0.822 & 0.847 & 269M \\
    \bottomrule
  \end{tabular}
  }}
\end{table}

\begin{figure}[htbp]
    \centering
    \includegraphics[width=0.8\linewidth]{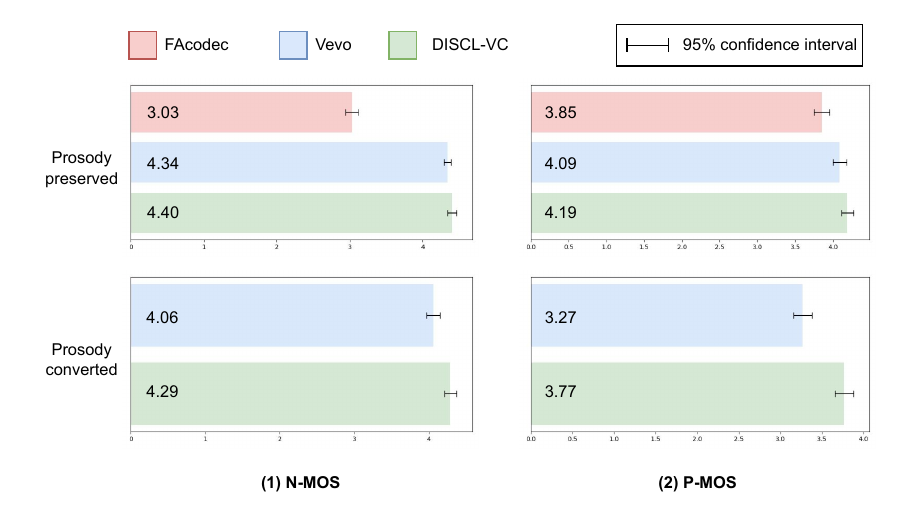}
    \caption{Results of subjective metrics for prosody related zero-shot voice conversion.}
    \label{fig:subjective}
\end{figure}

Table \ref{tab:esd} presents the objective metric results for prosody-related voice conversion experiments. For the prosody preservation task, we found that although FAcodec achieved better prosody similarity with the source speech, its speech naturalness and speaker similarity were significantly lower than those of other models. Meanwhile, our proposed Discl-VC outperforms Vevo in speech quality, prosody preservation, and speaker similarity, proving the effectiveness of our model. In the prosody conversion task, although our model exhibits lower speaker similarity compared to a baseline model with more than three times the number of parameters, it performs better across other evaluation metrics. This indicates that the proposed prosody mask transformer can generate prosody tokens of the source speaker based on the target speaker’s prosody, thus enabling effective prosody control. The results of subjective testing in Figure \ref{fig:subjective} support these findings as well.

\begin{figure}[htbp]
    \centering
    \includegraphics[width=0.7\linewidth]{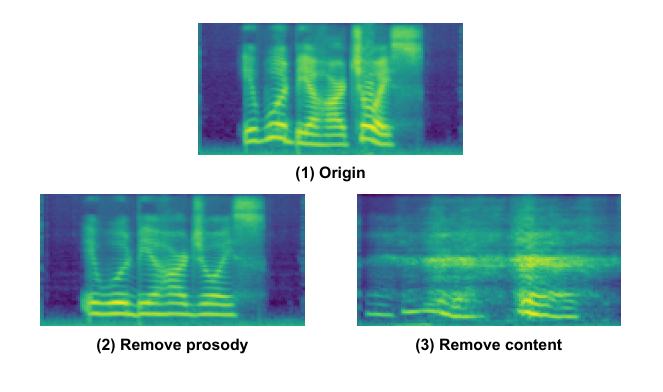}
    \caption{The visualization of speech disentanglement, where the corresponding sentence is: ``It would be a hard choice."}
    \label{fig:ablation}
\end{figure}

To validate the effectiveness of our model in disentangling prosody and content, we removed either the content tokens or prosody tokens and visually displayed the resulting speech via the mel spectrogram, as shown in the Figure \ref{fig:ablation}. It is clear that when prosody tokens is removed, the generated speech contains mostly pronunciation details, with almost all prosody information filtered out. On the other hand, when the content tokens are removed, the generated mel spectrogram lacks meaningful spoken content. This demonstrates that our model is capable of effectively disentangling prosody and content.

\subsection{Ablation study}

We also conducted ablation experiments on the VCTK dataset to validate the effectiveness of each component. Our experimental results, as shown in the Table \ref{tab:ablation}, demonstrate the following: when ContentVec is not used and the first 20 dimensions of the mel spectrogram \cite{Jiang2023MegaTTSZT} are employed as the input for prosody, the model's performance decreases across all metrics. We believe the key reason is that, although the first 20 dimensions of the mel spectrogram filter out some of the timbre and content information compared to the original mel spectrogram, during model training, since the source and target audio come from the same sample, the extracted prosody tokens inevitably still contain some speaker information, leading to mismatches between training and inference. When F0 loss is not used, the model performs better in terms of WER, but it shows weaker prosody similarity and other metrics compared to the original model. When using the original vector quantization instead of SimVQ, we also observed a decline in all objective metrics due to codebook collapse issues during training.

\begin{table}[htb]
  \caption{Ablation study.}
  \label{tab:ablation}
  \centering
    \scalebox{0.7}{
  \begin{tabular}{lcccc}
    \toprule
    \textbf{} & \textbf{UTMOS} ($\uparrow$) & 
    \textbf{WER} ($\downarrow$) & \textbf{F0\_Corr} ($\uparrow$) & \textbf{SECS} ($\uparrow$) \\
    \midrule
    Discl-VC & 4.079 & 1.946 & 0.973 & 0.929 \\
    \,\, w/o ContentVec & 4.034 & 3.147 & 0.971 & 0.900 \\
    \,\, w/o F0 loss & 4.048 & 1.279 & 0.971 & 0.928\\
    \,\, w/o SimVQ & 4.042 & 2.182 & 0.970 & 0.928 \\
    \bottomrule
  \end{tabular}
  }
\end{table}

\section{Conclusion}

In this paper, we propose a new controllable voice conversion framework that disentangles content, prosody, and timbre using different methods. And we leverage in-context learning to incorporate timbre information and generate high-fidelity speech with a flow matching transformer. Additionally, we introduce a non-autoregressive prosody mask transformer, which also predicts prosody tokens through in-context learning, enabling precise control over speech prosody. In future work, we plan to expand the model’s parameter size and train on larger datasets to further enhance model performance.

\section{Acknowledgements}

This work was supported in part by the National Natural Science Foundation of China under Grants 62276220 and 62371407 and the Innovation of Policing Science and Technology, Fujian province (Grant number: 2024Y0068)
 
\bibliographystyle{IEEEtran}
\bibliography{mybib}

\begin{thebibliography}{10}
\providecommand{\url}[1]{#1}
\csname url@samestyle\endcsname
\providecommand{\newblock}{\relax}
\providecommand{\bibinfo}[2]{#2}
\providecommand{\BIBentrySTDinterwordspacing}{\spaceskip=0pt\relax}
\providecommand{\BIBentryALTinterwordstretchfactor}{4}
\providecommand{\BIBentryALTinterwordspacing}{\spaceskip=\fontdimen2\font plus
\BIBentryALTinterwordstretchfactor\fontdimen3\font minus \fontdimen4\font\relax}
\providecommand{\BIBforeignlanguage}[2]{{%
\expandafter\ifx\csname l@#1\endcsname\relax
\typeout{** WARNING: IEEEtran.bst: No hyphenation pattern has been}%
\typeout{** loaded for the language `#1'. Using the pattern for}%
\typeout{** the default language instead.}%
\else
\language=\csname l@#1\endcsname
\fi
#2}}
\providecommand{\BIBdecl}{\relax}
\BIBdecl

\bibitem{Li2023SEFVCSE}
J.~Li, Y.~Guo, X.~Chen, and K.~Yu, ``Sef-vc: Speaker embedding free zero-shot voice conversion with cross attention,'' \emph{ICASSP 2024 - 2024 IEEE International Conference on Acoustics, Speech and Signal Processing (ICASSP)}, pp. 12\,296--12\,300, 2023.

\bibitem{ma24e_interspeech}
L.~Ma, X.~Zhu, Y.~Lv, Z.~Wang, Z.~Wang, W.~He, H.~Zhou, and L.~Xie, ``Vec-tok-vc+: Residual-enhanced robust zero-shot voice conversion with progressive constraints in a dual-mode training strategy,'' in \emph{Interspeech 2024}, 2024, pp. 2745--2749.

\bibitem{pmlr-v97-qian19c}
K.~Qian, Y.~Zhang, S.~Chang, X.~Yang, and M.~Hasegawa-Johnson, ``{A}uto{VC}: Zero-shot voice style transfer with only autoencoder loss,'' in \emph{Proceedings of the 36th International Conference on Machine Learning}, 2019, pp. 5210--5219.

\bibitem{pmlr-v119-qian20a}
K.~Qian, Y.~Zhang, S.~Chang, M.~Hasegawa-Johnson, and D.~Cox, ``Unsupervised speech decomposition via triple information bottleneck,'' in \emph{Proceedings of the 37th International Conference on Machine Learning}, 2020, pp. 7836--7846.

\bibitem{9747763}
C.~Ho~Chan, K.~Qian, Y.~Zhang, and M.~Hasegawa-Johnson, ``Speechsplit2.0: Unsupervised speech disentanglement for voice conversion without tuning autoencoder bottlenecks,'' in \emph{ICASSP 2022 - 2022 IEEE International Conference on Acoustics, Speech and Signal Processing (ICASSP)}, 2022, pp. 6332--6336.

\bibitem{Polyak2021SpeechRF}
A.~Polyak, Y.~Adi, J.~Copet, E.~Kharitonov, K.~Lakhotia, W.-N. Hsu, A.~rahman Mohamed, and E.~Dupoux, ``Speech resynthesis from discrete disentangled self-supervised representations,'' in \emph{Interspeech}, 2021.

\bibitem{pmlr-v162-qian22b}
K.~Qian, Y.~Zhang, H.~Gao, J.~Ni, C.-I. Lai, D.~Cox, M.~Hasegawa-Johnson, and S.~Chang, ``{C}ontent{V}ec: An improved self-supervised speech representation by disentangling speakers,'' in \emph{Proceedings of the 39th International Conference on Machine Learning}, 2022, pp. 18\,003--18\,017.

\bibitem{huang24_interspeech}
F.~Huang, K.~Zeng, and W.~Zhu, ``Diffvc+: Improving diffusion-based voice conversion for speaker anonymization,'' in \emph{Interspeech 2024}, 2024, pp. 4453--4457.

\bibitem{casanova2022yourtts}
E.~Casanova, J.~Weber, C.~D. Shulby, A.~C. Junior, E.~G{\"o}lge, and M.~A. Ponti, ``Yourtts: Towards zero-shot multi-speaker tts and zero-shot voice conversion for everyone,'' in \emph{International Conference on Machine Learning}, 2022, pp. 2709--2720.

\bibitem{popov2022diffusionbased}
V.~Popov, I.~Vovk, V.~Gogoryan, T.~Sadekova, M.~S. Kudinov, and J.~Wei, ``Diffusion-based voice conversion with fast maximum likelihood sampling scheme,'' in \emph{International Conference on Learning Representations}, 2022.

\bibitem{choi2024dddm}
H.-Y. Choi, S.-H. Lee, and S.-W. Lee, ``Dddm-vc: Decoupled denoising diffusion models with disentangled representation and prior mixup for verified robust voice conversion,'' in \emph{Proceedings of the AAAI Conference on Artificial Intelligence}, 2024, pp. 17\,862--17\,870.

\bibitem{Jiang2023MegaTTSZT}
Z.~Jiang, Y.~Ren, Z.~Ye, J.~Liu, C.~Zhang, Q.~Yang, S.~Ji, R.~Huang, C.~Wang, X.~Yin, Z.~Ma, and Z.~Zhao, ``Mega-tts: Zero-shot text-to-speech at scale with intrinsic inductive bias,'' \emph{ArXiv}, 2023.

\bibitem{jiang2024megatts}
Z.~Jiang, J.~Liu, Y.~Ren, J.~He, Z.~Ye, S.~Ji, Q.~Yang, C.~Zhang, P.~Wei, C.~Wang, X.~Yin, Z.~MA, and Z.~Zhao, ``Mega-{TTS} 2: Boosting prompting mechanisms for zero-shot speech synthesis,'' in \emph{The Twelfth International Conference on Learning Representations}, 2024.

\bibitem{10889108}
J.~Zuo, S.~Ji, M.~Fang, Z.~Jiang, X.~Cheng, Q.~Yang, W.~Liu, G.~Zhang, Z.~Tu, Y.~Guo, and Z.~Zhao, ``Enhancing expressive voice conversion with discrete pitch-conditioned flow matching model,'' in \emph{ICASSP 2025 - 2025 IEEE International Conference on Acoustics, Speech and Signal Processing (ICASSP)}, 2025.

\bibitem{Hussain2023ACEVCAA}
S.~S. Hussain, P.~Neekhara, J.~Huang, J.~Li, and B.~Ginsburg, ``Ace-vc: Adaptive and controllable voice conversion using explicitly disentangled self-supervised speech representations,'' \emph{ICASSP 2023 - 2023 IEEE International Conference on Acoustics, Speech and Signal Processing (ICASSP)}, pp. 1--5, 2023.

\bibitem{choi23d_interspeech}
H.-Y. Choi, S.-H. Lee, and S.-W. Lee, ``Diff-hiervc: Diffusion-based hierarchical voice conversion with robust pitch generation and masked prior for zero-shot speaker adaptation,'' in \emph{Interspeech 2023}, 2023, pp. 2283--2287.

\bibitem{Yao2024StableVCSC}
J.~Yao, Y.~Yang, Y.~Pan, Z.~Ning, J.~Ye, H.~Zhou, and L.~Xie, ``Stablevc: Style controllable zero-shot voice conversion with conditional flow matching,'' \emph{ArXiv}, 2024.

\bibitem{zhang2025vevo}
X.~Zhang, X.~Zhang, K.~Peng, Z.~Tang, V.~Manohar, Y.~Liu, J.~Hwang, D.~Li, Y.~Wang, J.~Chan, Y.~Huang, Z.~Wu, and M.~Ma, ``Vevo: Controllable zero-shot voice imitation with self-supervised disentanglement,'' in \emph{The Thirteenth International Conference on Learning Representations}, 2025.

\bibitem{lipman2023flow}
Y.~Lipman, R.~T.~Q. Chen, H.~Ben-Hamu, M.~Nickel, and M.~Le, ``Flow matching for generative modeling,'' in \emph{The Eleventh International Conference on Learning Representations}, 2023.

\bibitem{NEURIPS2023_2d8911db}
M.~Le, A.~Vyas, B.~Shi, B.~Karrer, L.~Sari, R.~Moritz, M.~Williamson, V.~Manohar, Y.~Adi, J.~Mahadeokar, and W.-N. Hsu, ``Voicebox: Text-guided multilingual universal speech generation at scale,'' in \emph{Advances in Neural Information Processing Systems}, 2023, pp. 14\,005--14\,034.

\bibitem{Guan2023ReflowTTSAR}
W.~Guan, Q.~Su, H.~Zhou, S.~Miao, X.~Xie, L.~Li, and Q.~Hong, ``Reflow-tts: A rectified flow model for high-fidelity text-to-speech,'' in \emph{ICASSP 2024 - 2024 IEEE International Conference on Acoustics, Speech and Signal Processing (ICASSP)}, 2024, pp. 10\,501--10\,505.

\bibitem{Guan2024LAFMAAL}
W.~Guan, K.~Wang, W.~Zhou, Y.~Wang, F.~Deng, H.~Wang, L.~Li, Q.~Hong, and Y.~Qin, ``Lafma: A latent flow matching model for text-to-audio generation,'' in \emph{Interspeech 2024}, 2024, pp. 4813--4817.

\bibitem{10889258}
K.~Wang, W.~Guan, S.~Lu, J.~Yao, L.~Li, and Q.~Hong, ``Slimspeech: Lightweight and efficient text-to-speech with slim rectified flow,'' in \emph{ICASSP 2025 - 2025 IEEE International Conference on Acoustics, Speech and Signal Processing (ICASSP)}, 2025, pp. 1--5.

\bibitem{10832320}
S.~E. Eskimez, X.~Wang, M.~Thakker, C.~Li, C.-H. Tsai, Z.~Xiao, H.~Yang, Z.~Zhu, M.~Tang, X.~Tan, Y.~Liu, S.~Zhao, and N.~Kanda, ``E2 tts: Embarrassingly easy fully non-autoregressive zero-shot tts,'' in \emph{2024 IEEE Spoken Language Technology Workshop (SLT)}, 2024, pp. 682--689.

\bibitem{chen-etal-2024-f5tts}
Y.~Chen, Z.~Niu, Z.~Ma, K.~Deng, C.~Wang, J.~Zhao, K.~Yu, and X.~Chen, ``F5-tts: A fairytaler that fakes fluent and faithful speech with flow matching,'' \emph{arXiv preprint arXiv:2410.06885}, 2024.

\bibitem{chang2022maskgit}
H.~Chang, H.~Zhang, L.~Jiang, C.~Liu, and W.~T. Freeman, ``Maskgit: Masked generative image transformer,'' in \emph{The IEEE Conference on Computer Vision and Pattern Recognition (CVPR)}, 2022.

\bibitem{ju2024naturalspeech}
Z.~Ju, Y.~Wang, K.~Shen, X.~Tan, D.~Xin, D.~Yang, E.~Liu, Y.~Leng, K.~Song, S.~Tang, Z.~Wu, T.~Qin, X.~Li, W.~Ye, S.~Zhang, J.~Bian, L.~He, J.~Li, and sheng zhao, ``Naturalspeech 3: Zero-shot speech synthesis with factorized codec and diffusion models,'' in \emph{Forty-first International Conference on Machine Learning}, 2024.

\bibitem{wang2025maskgct}
Y.~Wang, H.~Zhan, L.~Liu, R.~Zeng, H.~Guo, J.~Zheng, Q.~Zhang, X.~Zhang, S.~Zhang, and Z.~Wu, ``Mask{GCT}: Zero-shot text-to-speech with masked generative codec transformer,'' in \emph{The Thirteenth International Conference on Learning Representations}, 2025.

\bibitem{9585401}
W.-N. Hsu, B.~Bolte, Y.-H.~H. Tsai, K.~Lakhotia, R.~Salakhutdinov, and A.~Mohamed, ``Hubert: Self-supervised speech representation learning by masked prediction of hidden units,'' \emph{IEEE/ACM Transactions on Audio, Speech, and Language Processing}, pp. 3451--3460, 2021.

\bibitem{Zhu2024AddressingRC}
Y.~Zhu, B.~Li, Y.~Xin, and L.~Xu, ``Addressing representation collapse in vector quantized models with one linear layer,'' \emph{ArXiv}, 2024.

\bibitem{ren2021fastspeech}
Y.~Ren, C.~Hu, X.~Tan, T.~Qin, S.~Zhao, Z.~Zhao, and T.-Y. Liu, ``Fastspeech 2: Fast and high-quality end-to-end text to speech,'' in \emph{International Conference on Learning Representations}, 2021.

\bibitem{Borsos2023SoundStormEP}
Z.~Borsos, M.~Sharifi, D.~Vincent, E.~Kharitonov, N.~Zeghidour, and M.~Tagliasacchi, ``Soundstorm: Efficient parallel audio generation,'' \emph{ArXiv}, 2023.

\bibitem{lee2023bigvgan}
S.~gil Lee, W.~Ping, B.~Ginsburg, B.~Catanzaro, and S.~Yoon, ``Big{VGAN}: A universal neural vocoder with large-scale training,'' in \emph{The Eleventh International Conference on Learning Representations}, 2023.

\end{thebibliography}

\end{document}